\documentclass[useAMS,usenatbib]{mnras}
\usepackage{amsfonts}
\usepackage{amsmath}
\usepackage{amssymb}
\usepackage{subcaption}
\usepackage[dvips]{graphicx}
\usepackage{color}
\usepackage{multirow}
\def\bea{\begin{eqnarray}}
\def\ena{\end{eqnarray}}

\newcommand{\mr}[1]{\mathrm{#1}}

\definecolor{blazeorange}{rgb}{1.0, 0.4, 0.0}

\title[Gamma-ray flux depressions of the Crab nebula in the high-energy range]
{Gamma-ray flux depressions of the Crab nebula in the high-energy range}

\author[M. S. Pshirkov et al.]
{M. S. Pshirkov$^{1,2,3}$\thanks{E-mail:pshirkov@sai.msu.ru (MSP)},
 B. A. Nizamov $^{1}$\thanks{E-mail: nizamov@physics.msu.ru (BAN)},
A. M. Bykov $^{4}$\thanks{E-mail: byk@astro.ioffe.ru  (AMB)}
and  Yu. A. Uvarov$^{4}$\thanks{E-mail:uv@astro.ioffe.ru
 (YAU)}\\
$^{1}$
Sternberg Astronomical Institute, Moscow State University, 119992, 
Universitetskiy Prospekt, 13, Moscow, Russia\\
$^{2}$
Institute for Nuclear Research of the Russian Academy of Sciences, 117312, Moscow, Russia\\
$^{3}$
P. N. Lebedev Physical Institute of the Russian Academy of Sciences, Pushchino Radio Astronomy Observatory,
Pushchino 142290, Russia\\
$^{4}$ Ioffe Institute, 26 Politekhnicheskaya st., St. Petersburg 194021, Russia\\
}

\begin{document}

\date{}

\pubyear{2020}

\maketitle

\label{firstpage}

\begin{abstract}
The giant  gamma-ray flares of the Crab nebula discovered by
AGILE and {\em Fermi} observatories came as a surprise and have challenged the
existing models of pulsar wind nebulae. We have carried out an analysis
of 10.5 years of {\em Fermi}-LAT observations (Aug 2008 -- Feb 2019) and 
investigated variability of the Crab nebula in the 100-300 MeV range. 
Besides the flares, we found several month long depressions
of the gamma-ray flux and identified several cases of sharp flux drops,
where during one week the flux decreased by an order of magnitude with respect to its average value. No statistically significant variations of
the nebula flux in the $E>$10~GeV range were found in the data. We
discuss possible implications of the observed gamma-ray flux depressions 
on the model of synchrotron emission of the Crab nebula. 
\end{abstract}

\begin{keywords}
ISM: supernova remnants -- gamma-rays -- supernovae: individual (Crab nebula)
\end{keywords}

\section{Introduction}
\label{sec:intro}
The Crab nebula and the pulsar at the center of this nebula are 
among the most important objects of modern astrophysics 
\citep[][]{2008ARA&A..46..127H}. Given its role as the primary test bench for
studies of a broad class of pulsar wind nebulae (PWNe) and young pulsars, the Crab 
nebula have been observed with high cadence across the whole electromagnetic spectrum. 
Almost since its discovery in the high-energy range, the Crab nebula has been 
considered a perfect calibration source for many astronomical observations. Due
to its spatial extent it was assumed that its flux in the X-rays and at higher energies should be rather stable.
However, over the last 15 years observations revealed that this assumption
is not always correct \citep{BB14}: first, {\em Fermi}/GBM observations in the
15-50~keV range showed a $\sim7\%$ flux decrease in a 2 years time span (2008-2010);
this result was corroborated by observations with other instruments, such as {\em
Swift}/BAT, {\em RXTE}/PCA, {\em INTEGRAL}/IBIS \citep{Wilson-Hodge2011}.
Even more striking and unexpected was the discovery of flaring emission 
in the 0.1-0.5~GeV range made with {\em AGILE} and {\em Fermi}-LAT \citep{Tavani2011,Abdo2011a}. 
The brightest flare detected to date occurred in April 2011
and lasted approximately 9 days, the peak flux above 100 MeV increased
30-fold compared to the average value \citep{Buehler2012}. 
The spectrum of gamma-ray emission changed drastically during the flares. 
During the steady state it is well reproduced by a combination of 
two components (see Fig. \ref{fig:spectrum}), the first one being a power law with a very soft spectral 
index $\gamma_s\sim3.6$, and the second one -- a broken power law 
with the spectral index gradually softening from $\gamma_{I,1}\sim1.5$ 
at lower energies to $\gamma_{I,2}\sim2.2$ at energies higher than 10~GeV. 
The first component is usually attributed to the synchrotron emission 
from short-lived PeV range electrons, while the latter is thought to be
produced by an inverse Compton (IC) emission from a population of multi-GeV 
electrons.
During the flares, the amplitude of the synchrotron component increased
and also the spectral index became much harder, $\gamma_s\sim1.3$
with emerging exponential cut-off at around 300~MeV. A summary of 
the flaring activity of the Crab nebula observed with {\em Fermi}-LAT 
for the last 11 years has been recently compiled by \citet{Huang2020}.

The bulk emission of the Crab nebula from the radio band to the sub-GeV 
gamma-ray range is mainly due to the synchrotron radiation of relativistic
electrons and positrons \citep[see, e.g.,][]{1996MNRAS.278..525A}. However,
explanation of the GeV regime flares of the nebula with a synchrotron 
model is non-trivial. The fastest acceleration time of an electron in 
an ideal MHD flow with the frozen-in magnetic field should not be 
substantially shorter than a particle gyro-period. Hence the energy $E_m$ 
of a synchrotron photon emitted by an electron or positron of the maximal 
energy in an accelerator with a steady magnitude of the root mean square (r.m.s.)  magnetic field 
where the synchrotron cooling rate is balanced by the particle
acceleration rate would be less than $\approx$ 30~MeV (independent
on the value of the magnetic field) \citep{Arons12}. It is difficult to overcome this limit in quasi-steady and statistically homogeneous systems. However, if electrons are accelerated at the wind termination surface with some r.m.s. magnetic field but radiate in higher field (in a transient regime) 
their synchrotron photons would have energies well above $E_m$. A flare can be produced by a filament of much higher magnetic field inflowing into the accelerator. 
The magnitude of the flux change depends on the magnetic field variation 
and it is highly amplified in the cut-off region of the electron 
spectrum \citep{Bykovea12}. A flare magnitude can be high in the sub-GeV synchrotron 
range while appearing very modest at lower synchrotron photon energies. Fast cooling in the high magnetic field filament then require some time to restore a population of the highest energy particles which is accompanied with  photon flux depression.  
There are also models of the gamma-ray flares 
observed in the Crab nebula, based on pair acceleration by electric 
fields in magnetic reconnection regions, developed by \citet{Ceruttiea13}. 
Doppler boosting of the photons produced at energies below $E_m$ 
in the vicinity of the nebula's termination shock was considered 
in the model by \citet{KL11}. Discussions of various aspects 
of gamma-ray flare modeling can be also found in 
\citep{BB14,SKL15,Lemoine16,Zrake16,Wernerea16,Porthea17,SL20,Pohlea20}. 
A thorough analysis of the high energy variability of the Crab nebula 
is needed to reveal the physical nature of the puzzling flare phenomenon.

In this paper we have carried out an analysis of the high-energy variability
of the Crab nebula through a more than 10 year time span (Aug 2008 -- Feb
2019), primarily focusing at the 100-300 MeV range emission where the
synchrotron component is prominent.

\section{Data and methods}
\label{sec:data}
\begin{figure}
\begin{center}
\includegraphics[angle=270, width=\columnwidth ]{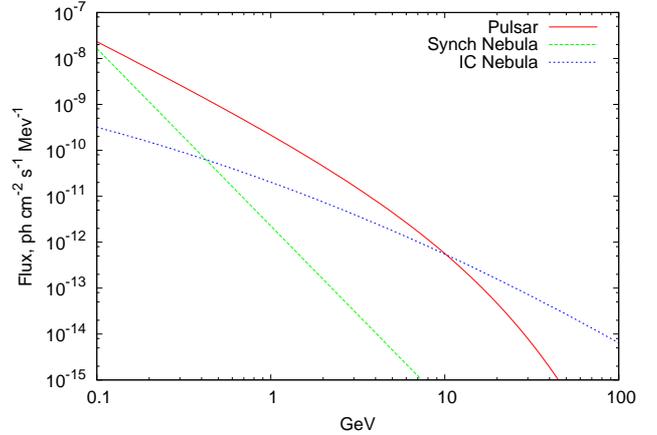}
\end{center}
\caption{Spectrum of the Crab Nebula and the Crab pulsar based on the 4FGL catalogue spectral models.} \label{fig:spectrum}
\end{figure}

The analysis of the Crab nebula is strongly affected by a local background provided by a very bright Crab pulsar. In the latest 8-year source catalog 4FGL these sources are modelled using 3 independent components -- nebula synchrotron and IC and pulsar proper \citep{4FGL}. The spectrum is shown in the Fig. \ref{fig:spectrum}. It can be seen that:
1) The flux from the pulsar completely dominates at energies above 100 MeV but below 10 GeV  2) the synchrotron component is considerably larger than the  IC component in 100-300 MeV energy range 3) at energies higher than 10 GeV the most of the total flux is produced by the IC component.

It is evident that the investigation of the synchrotron component should be made in 100-300 MeV range, otherwise there would be strong confusion from the IC component.

In our analysis we have used 130 months of Fermi LAT data collected  since 2008 Aug 04 ( MET =239557417 s)  
until 2019 Mar 01 (MET=573091205).  We have selected events that belong to the "SOURCE" 
class. The Pass 8R3 reconstruction and {\em Fermitools} \footnote{http://fermi.gsfc.nasa.gov/ssc/data/analysis/software/}   were used. A usual event quality cut, namely that the zenith angle should be   less than $90^{\circ}$    was imposed.

We took a circle of 15$^{\circ}$ radius around the position of the Crab pulsar
($\alpha_{J2000}=83^{\circ}.6331, \delta_{J2000}=22^{\circ}.0145$) as our region of interest (RoI). 

\begin{figure}
\begin{center}
\includegraphics[angle=270, width=\columnwidth ]{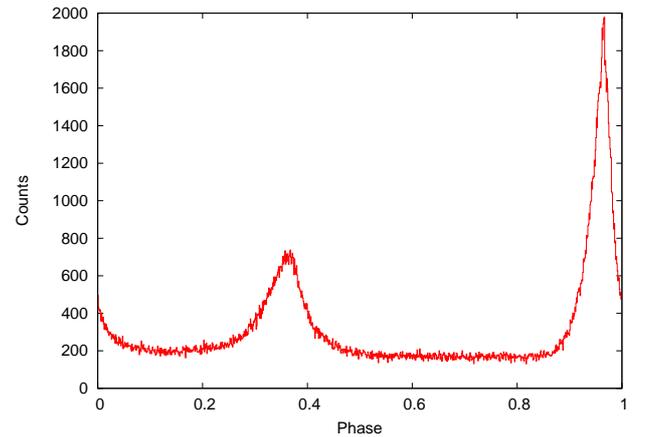}
\end{center}
\caption{ Phase profile obtained from gamma rays with energies above 100 MeV within $3^{\circ}$ of the Crab pulsar (10.5 years of observations, 318255 events).} \label{fig:pulse}
\end{figure}

It is absolutely necessary to suppress the contribution from the pulsar, and fortunately it can be done because the pulse profile demonstrates large off-pulse region where the pulsar emission is essentially absent \citep{Abdo2011a}. For further studies we have defined $0.5<\phi<0.85$ as our off-pulse region, while the main peak maximum is located  at $\phi_{\mr{max}}=0.965$ (see Fig. \ref{fig:pulse})

Each photon was assigned phase using {\em gtpphase} procedure of the {\em Fermitools} package. Time of arrival (ToA) of photon  and exact position of the satellite at that moment allow to calculate barycentric ToA. Pulsar rotational ephemeris --  its frequency and its derivatives were needed to get pulse phases from the ToAs.
As the Crab pulsar is very young, its rotation is highly irregular, sometimes it even experiences so-called glitches -- rapid increases of rotational frequency. Because of that it is extremely difficult to construct ephemeris even for one year time span, let alone for more than decade. To overcome this difficulty we have used Jodrell Bank Crab pulsar monthly ephemeris \citep{Lyne1993} \footnote{ http://www.jb.man.ac.uk/~pulsar/crab.html}. The data were split into segments according to the ephemeris bins, most of these segments coincide with calendar months. After that ToAs of photons in each  segment were folded with corresponding ephemeris and  phase was assigned to each photon.

We performed analysis in 100-300 MeV energy range  for three phase regions: off-pulse, on-pulse and full pulse. Light curves were obtained using two different types of binning -- first, we used one-month bin. Also, due to the brightness of the source, it is possible to go even further and use smaller one-week bins. Even  smaller bins, e.g. half-weeks would lack statistics during the strongest depressions when the flux of the nebula drops to its minimal values. 

We used standard {\em gtlike} utility for the analysis. The  source model included 184  sources from the 4FGL catalogue \citep{4FGL}, contributing inside the RoI, the latest galactic interstellar emission model gll$\_$iem$\_$v07.fits, and the isotropic spectral template iso$\_$P8R3$\_$SOURCE$\_$V2$\_$v1.txt\footnote{http://fermi.gsfc.nasa.gov/ssc/data/\\access/lat/BackgroundModels.html}. 
We parametrized the contribution from the Crab with a simple power-law  function with all parameters left free to vary. 
Apart from that only normalizations of the backgrounds were left free in the fitting procedure, all  parameters of other sources were set to their values given in the 4FGL  catalogue. The contribution from the Sun was studied and it was less than the errors of the fit  even for weekly binning.

\section{Results and discussion}
\label{sec:discussion}
\begin{figure}
\begin{center}
\includegraphics[angle=270, width=\columnwidth ]{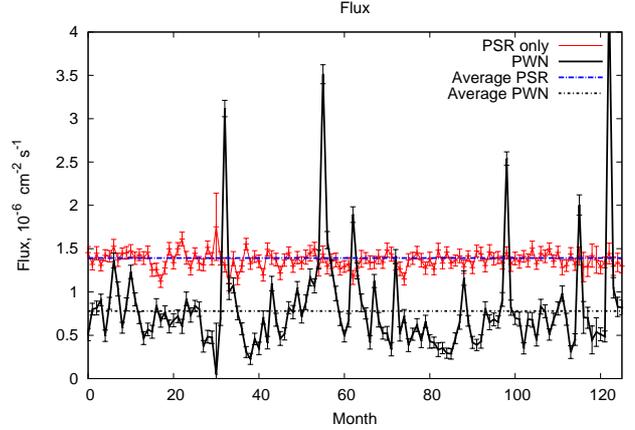}
\end{center}
\caption{ Light curve of the Crab nebula and  the Crab pulsar, one-month bin. } \label{fig:lc_month}
\end{figure}

\begin{figure}
\begin{center}
\includegraphics[angle=270, width=\columnwidth ]{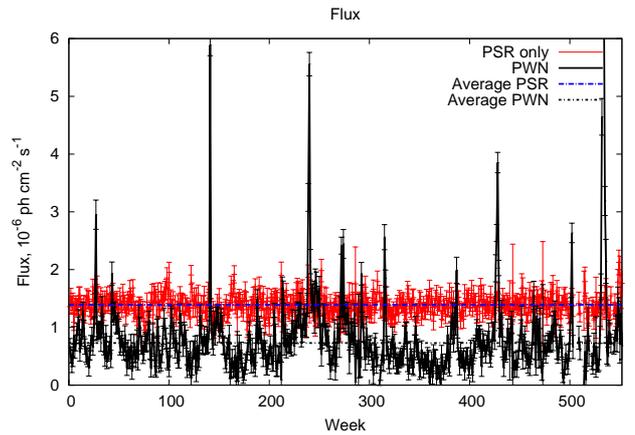}
\end{center}
\caption{  Light curve of the Crab nebula and  the Crab pulsar, one-week bin.} \label{fig:lc_week}
\end{figure}

The main results of the analysis are presented in Figs. \ref{fig:lc_month} and \ref{fig:lc_week}. 
Flux  of the pulsar $F_{PSR}$ in each bin was calculated as $F_{PSR}=F_{on}-\frac{1-\Delta_{off}}{\Delta_{off}}F_{off}$,  where $F_{on},F_{off}$ are fluxes in on-pulse and off-pulse regions, $\Delta_{off}=0.35$ is the phase duration of the off-pulse part. Similarly, flux of the nebula was calculated as $F_{PWN}=F_{off}/\Delta_{off}$.

\begin{figure}
\begin{center}
\includegraphics[angle=270, width=\columnwidth ]{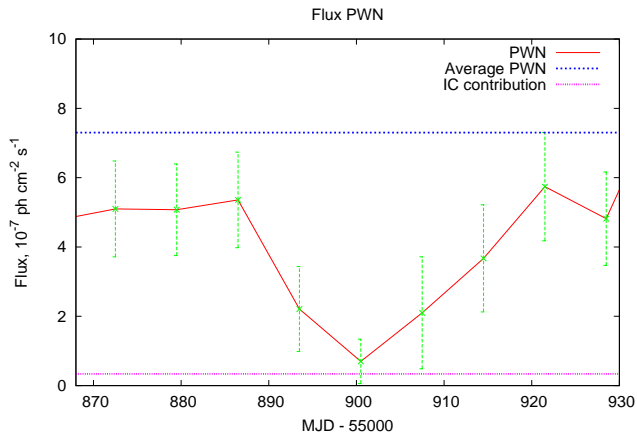}
\end{center}
\caption{  Light curve of the Crab nebula Oct 2011- Jan 2012.} \label{fig:lc_week_201112}
\end{figure}

Several strong flares are the most prominent features of the light curves at both binning scales. The pulsar emission was very stable during the considered time period. It was definitely not the case for the  component coming from the nebula -- apart from flares it demonstrated long rises and depressions. This behaviour was earlier found in the analyses of {\em EGRET}, {\em AGILE} and Fermi LAT (up to May 2012) data as well \citep{deJager1996,Striani2013}. In the latter analysis \citep{Striani2013} no specific selection of off-pulse photons was  performed.
In what follows we focus our analysis on the 'depression' episodes. The most prominent one had place during December 2011, and another strong depressions were observed in June and August 2015 and January 2018.
Week light curve of the Oct 2011 - Jan 2012 interval  is shown in Fig. \ref{fig:lc_week_201112}. During this interval the PWN flux was considerably below its average value $\bar{F}_{PWN}=7.3\times10^{-7}~\mr{ph~cm^{-2}~s^{-1}}$ which was calculated using all epochs in absence of strong flares.  But the most striking feature is a strong dip around MJD 55900 (first week of December 2011). In this bin the flux decreased to $F_{min}=(7\pm6)\times10^{-8}~\mr{ph~cm^{-2}~s^{-1}}$ which is order of magnitude smaller than the average value and three times smaller than the flux in the previous bin. There are three additional  examples of such extreme variability (epochs: MJD 57197, 57244, 58128) where the flux falls to comparable values.  It can be seen that in the minimum the level of the nebula flux is  very close to the level of contribution from the IC component (see Fig. \ref{fig:spectrum}), assumed to be non-variable.

\begin{figure}
\centering
\includegraphics[angle=0, width=\columnwidth ]{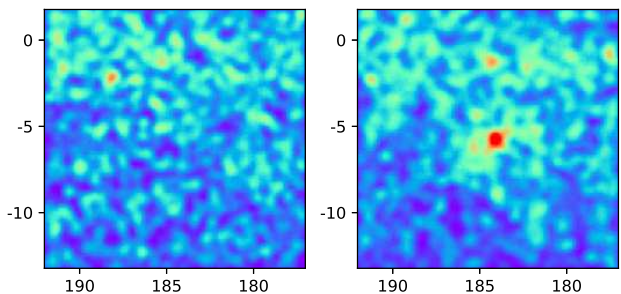} \\

\caption{\textit{Left panel}: count map for 4 weeks with extremely low flux. \textit{Right panel}: count map for 4 weeks with flux close to its average value.} \label{fig:cmap}
\end{figure}

It can be demonstrated  that it is not an artefact of the fit. We combined count maps for 4 weeks with extremely low flux and 4 weeks where the flux was close to its average value (Fig. \ref{fig:cmap}). In the first case the nebula is almost absent from the count map and the most prominent features belong to diffuse background.  In the second case the very same features are clearly subdominant comparing to the emission from the nebula. The source  in the first case were detected with very low significance, test statistics $TS\sim2,~F_{PWN}=(6.5\pm6.3)\times10^{-8}~\mr{cm^{-2}~s^{-1}}$, in sharp contrast with  the second case:   $TS\sim160,~F_{PWN}=(8.0\pm0.9)\times10^{-7}~\mr{cm^{-2}~s^{-1}}$.

Also, we have studied the behaviour of the source at  high energies, $E>10~$GeV, where the hard, presumably IC component from the nebula starts to dominate over the pulsar. Thus it is possible to use full phase window without losing almost two thirds of information because of application of gating procedure. Flux is considerably lower at these energies and sheer lack of statistics does not allow to obtain meaningful results with binning finer than one month when performing maximum likelihood analysis with the {\em gtlike} utility.
However, much improved angular resolution at these energies makes possible robust check with means of aperture photometry -- we took selection radius equal to $0^{\circ}.5$ which is close to the value of $PSF_{95}$  (point-spread function which contains 95\% of all photons from the source). The Crab nebula is by far the brightest source at energies $E>10$~GeV so we operated in virtually background-free regime. Photon counts and exposures were calculated for two types of  bins (month, week) which allowed to calculate expected number of counts in bins.  Average flux was set equal to ratio of the  total numbers of photons to total exposure. After that the observations were compared to expectations -- we calculated local p-values, assuming that in every bin the number of observed photons followed the Poissonian distribution. No significant fluctuations were found -- in case of weekly binning the lowest local p-value was equal to $1.2\times10^{-3}$ which  after application of the 'look elsewhere' correction gave the global p-value around 0.6, the lowest global p-value for one-month binning was equal to 0.3.

\begin{figure}
\centering
\includegraphics[angle=0, width=\columnwidth ]{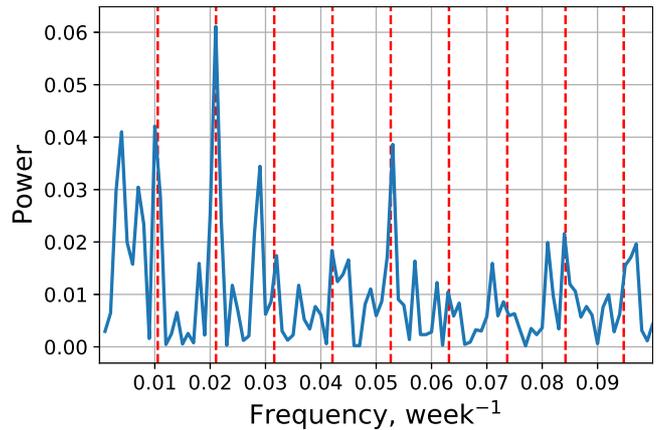} \\

\caption{Lomb-Scargle peridogram of the light curve. Dashed horizontal lines indicate harmonics of the fundamental period $P_0=95$~weeks.} \label{fig:periodogram}
\end{figure}

We have tried to detect any periodicity in the observed  light curve. We have calculated  Lomb-Scargle periodogram \citep{Lomb1976,Scargle1982}, using one-week binning.  We have find main peak at the period corresponding to $P_0\sim47.5$~weeks. As there were also additional weaker peaks corresponding to periods $2P_0$, $P_0/1.5$, $P_0/2$, $P_0/2.5$ and  even higher harmonics (see Fig. \ref{fig:periodogram}), it could indicate a presence of fundamental period at $\sim95$~weeks.

Finally, we have performed a search for a possible asymmetry in the rise and fall timescales both for the full light curve and for the light curve in the immediate vicinity of significant depressions\footnote{We thank the referee for suggesting this test.} and could not find any statistically significant deviations from a symmetry.

We have tried to interpret the observed depressions in the model by \citet{Bykovea12} where the pulsar wind in the upstream flow is inhomogeneous and can fluctuate by factor several comparing to its mean value. The high energy electrons which are responsible for the synchrotron emission in the 100 MeV energy band are concentrated in a narrow layer with the width of $L_{emm,cm}=1.6\times10^{3}\frac{U_{d,cm/s}}{H_{G}^{3/2} \nu_{keV}^{1/2}}$
near the termination shock. For the downstream speed $U_{d,cm/s}=c/3$
and magnetic field $H_{G}=2\times10^{-4}$~G this is $L_{emm}=2\times10^{16}$~cm which corresponds to the characteristic time of the flux change equal to $\tau\sim2\times10^{6}$~s. If the electron concentration meets a region with enhanced magnetic field, a flare occurs. In opposite case, when the magnetic field is temporarily  lower  than its average value, there is a decrease in the synchrotron emission. The termination shock region is an extended object and the time delay of photons that come to the observer
from different parts smears the light curve of the emission. It could be not too important for a flaring case, when depending on a degree of the magnetic field enhancement, the 'hot-spot' could overshine the rest of the emission layer, but it strongly affects the opposite case of flux depression.

Based on Chandra data and distance estimation of 2 kpc we assume the
termination shock radius to be $4\times10^{17}$cm and the pulsar equatorial
plane inclination angle to the line of sight to be $\sim30^{\circ}$. 

\begin{figure}
\includegraphics[scale=0.5, angle=0]{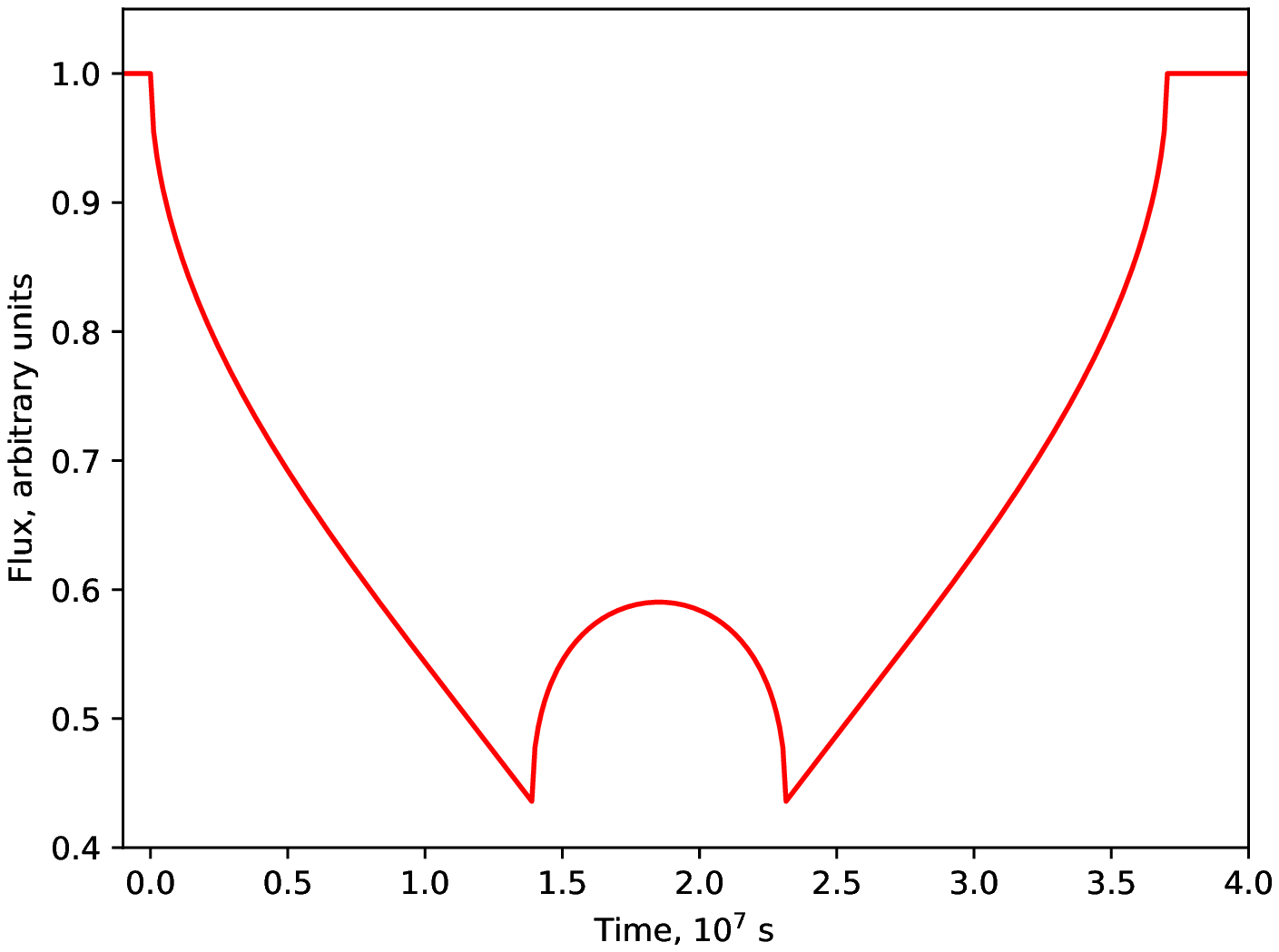}\\
\includegraphics[scale=0.5, angle=0]{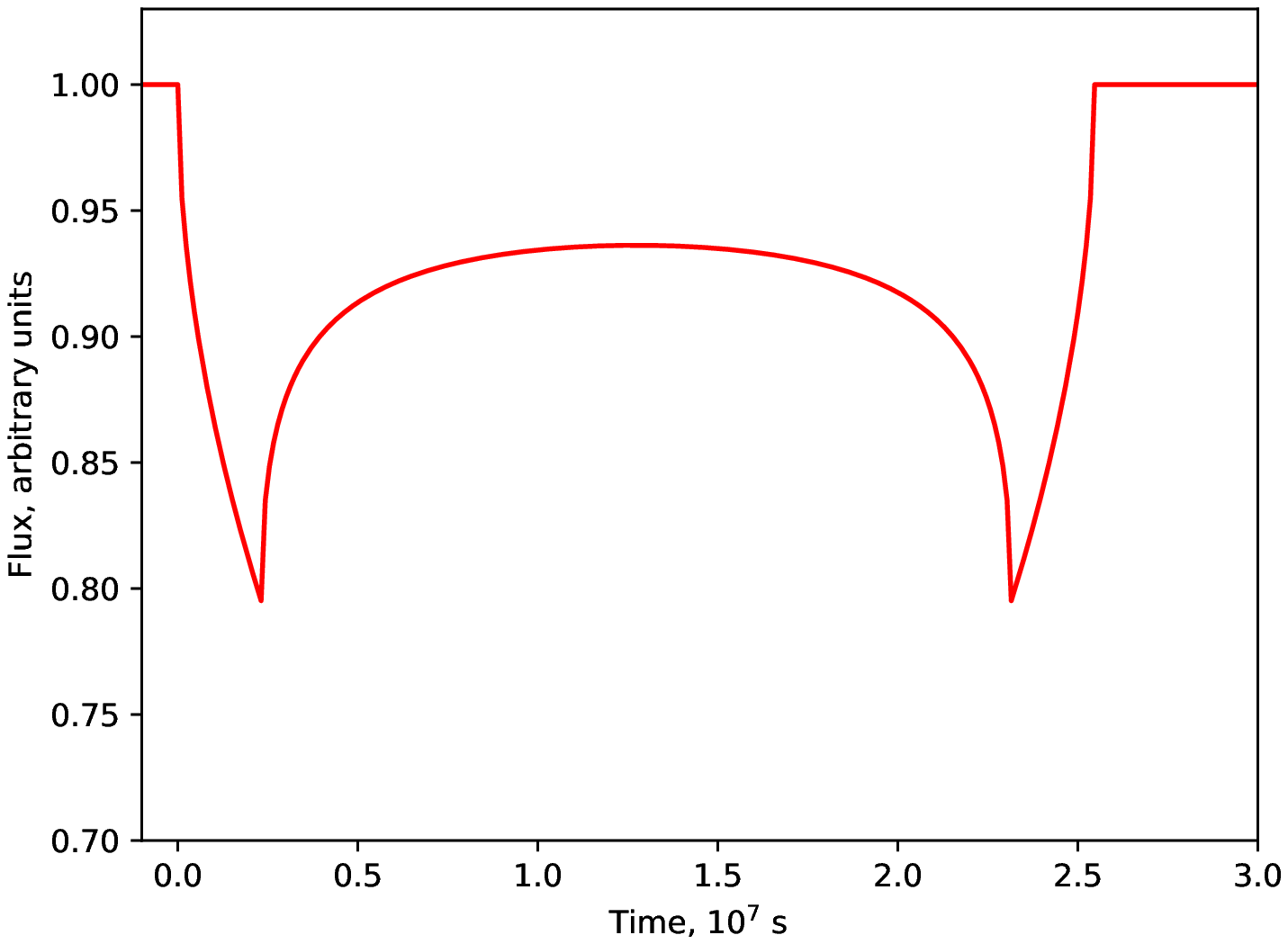}\\
\includegraphics[scale=0.5, angle=0]{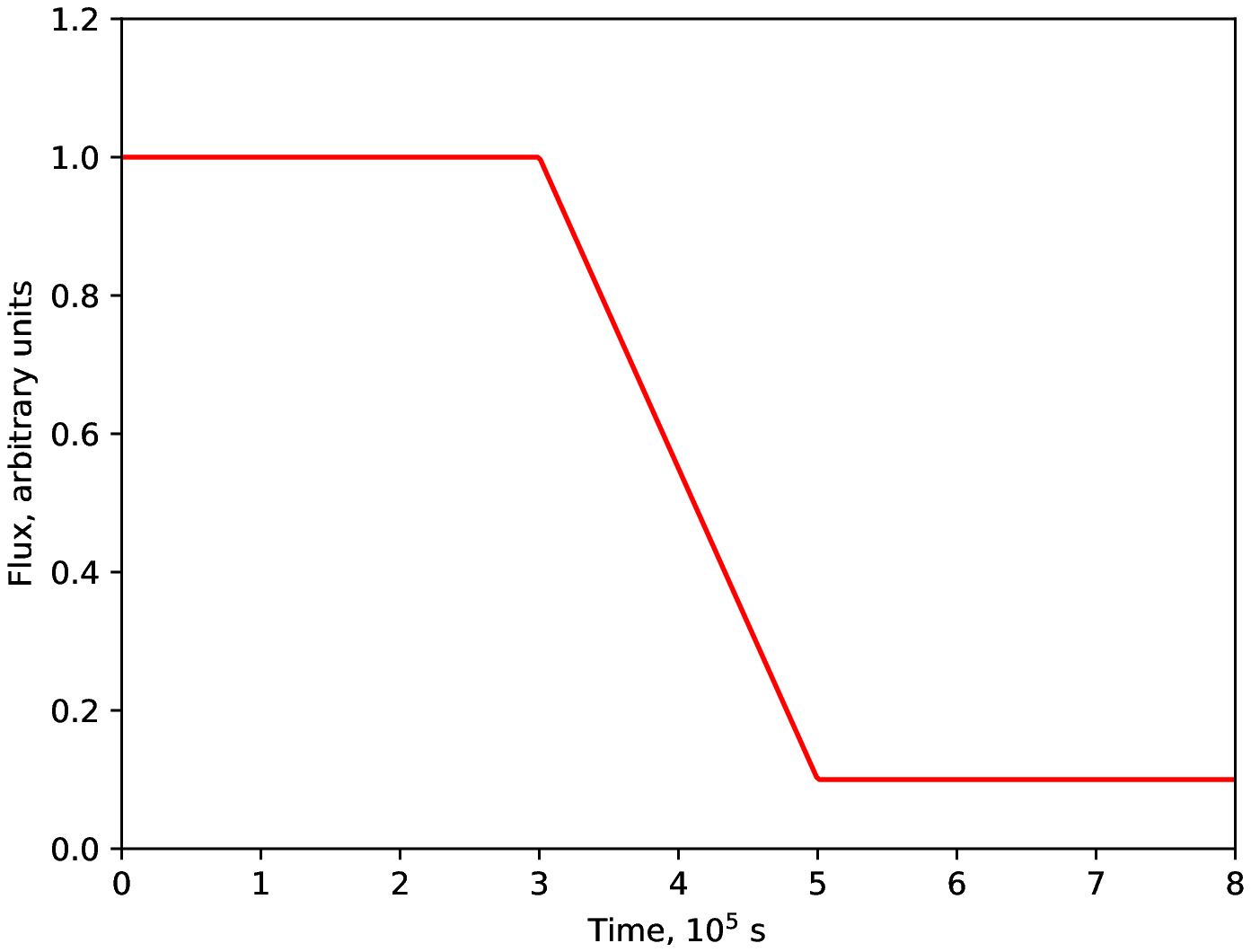}

\caption{This figure shows simulated light curves produced by a magnetic field inhomogeneity  incident on the particle acceleration region downstream of the wind termination surface.  
The top and middle panels illustrate the simulated light curves produced by the "magnetic wells" of $1.4\times10^{7}$s (top) and $2.5\times10^{6}$s (middle) duration in the non-relativistic downstream outflow. Here we assumed the model of axially symmetric PWN with most of the emission coming from the equatorial part of the termination surface. The bottom panel illustrates the possible origin of sharp edges (both drops or rises) produced by a similar fluctuation but in the relativistic downstream flow with $\Gamma_d \sim 3$ of an oblique part of the termination surface. In this case the emission comes from the compact region
where the plasma flow is directed towards the observer so emission region geometry is almost
plain and does not affects the lightcurve.    \label{fig:lightcurve}}

\end{figure}

We consider a model of axially symmetric pulsar wind nebula like that presented by \citet{KL11} and \citet{Porth14} assuming that the quasi-steady electron distribution is formed in the vicinity of the pulsar wind termination surface. The gamma-ray  100-300 MeV photons are produced by the synchrotron radiation of the accelerated PeV regime pairs. The time variation of the gamma-ray flux can be initiated by magnetic field fluctuations incident in the region filled with the accelerated pairs.   
In the 100-300 MeV energy range the flux would fall by 10 times if the 
incoming magnetic field falls by 2 times. 
However, due to the time delay we can still see the
emission from the farther parts of the termination shock during the time
$\tau\sim$2.3$\times$10$^{7}$s. In this  model, the observable light curve is also sensitive to the duration of the depression of the magnetic field  in the wind. If "magnetic well" duration is less than the light travel time
across the nebula then the emission never totally turns off and the observable depression amplitude is controlled by the relation between these times and system geometry. 
Top and middle panels of the Figure~\ref{fig:lightcurve} 
show simulated light curves for two different assumptions  on the magnetic field decrease duration. In both cases the time duration of the "magnetic wells" is less than the light travel time
across the nebula that results in the formation of a central bump at the lightcurves. The bump formation begins at a time when the magnetic field depression  leaves the  termination shock and flux from the part of the TS nearest to the observer is restored to the pre-depression level. It ends at a time when the decreased  radiation due to magnetic field depression from the farthest part of the TS reaches the observer.

Whole  emission  at the energies larger than 100 MeV is concentrated in a thin layer with $2\times10^{16}$~cm thickness which can be spatially resolved by the Chandra X-ray observatory. 
The expected variation of the observed X-ray flux from this layer should be low, because the emission  in a $\mathcal{O}$(keV) energy 
range contains a substantial background fraction accumulated  along the line of sight from the much wider region of the nebula, precluding large variations in the Chandra data.

The model of a filament in non-relativistic flow could explain both flares and flux depressions at longer time scales (several month) but can not explain sharp drops of flux, like observed in December 2011, when the flux decreased by a factor of several at $\mathcal{O}(10^{6}$~s) timescales. If instead the relativistic upstream flow produces
the emission or if the termination shock is
oblique and the downstream flow is relativistic, then different parts
of the emitting surface have different Doppler factors and the bulk of
the emission comes from the regions moving in the direction to the observer within angles $\sim 1/\Gamma_d$, where $\Gamma_d$ is a Lorentz factor of the relativistic flow. In
this case the characteristic variability time 
is determined not by the time delay of the photons coming
from the whole layer, but only by those coming from a small region 
where plasma flows towards the observer. Even more, if $L$ is the size of the emission region along the line of sight and $L/c$ is the time needed for relativistic flow  to pass through this region, the observer will see this process during shorter time of $L/(2\Gamma_d^2c)$. The size and form of this emission region are determined in this case by the geometry of the plasma flow lines with respect to the direction to observer. The MHD modeling of PWN is beyond of the scope of this article but we made estimations using the results of 3D MHD model of the Crab nebula by \citet{Porth14}. In this model the size  of the emission region coincident with a part of the oblique termination shock is $L\sim6\times10^{16}$ cm. In the very simplified case of the plane emitting region (shown in the bottom panel of Fig.~\ref{fig:lightcurve}) the transition in the light curve from  the  high level of the  emission in 100-300 MeV energy band  to the low (and vice versa) occurs in $2\times10^5$s if $\Gamma_d\approx2.2$. For more realistic geometry the same result would be obtained with $\Gamma_d \approx3.5$ which is still a realistic value of $\Gamma_d$ (see e.g. \citet{KL11}). Moreover, the model by \citet{Porth14} demonstrated a possible presence of the quasi-periodic distortions of the termination shock with the characteristic periods consistent with that seen in Fig.~\ref{fig:periodogram}. The observed time variability of the gamma-ray flux above 100 MeV will certainly help to construct and test future more realistic models of the Crab nebula.  

\section{Conclusions}
\label{sec:conclusions}
With 10.5 years of the {\em Fermi}-LAT observations we have investigated 
variability of emission from the Crab nebula in the 100-300~MeV range.
Besides already well known flares we have found several month long flux depressions 
and identified several cases of sharp flux drops, where during one week 
the gamma-ray flux decreased by an order of magnitude with respect to its average 
value and by the factor of three compared to the previous week. 
However, no significant variations in the $E>10$~GeV energy range were found. 
This picture is consistent with the hypothesis that the emission above GeV 
is dominated by the inverse-Compton 
scattered photons from a quasi-steady accelerator of PeV energy regime electron-positron
pairs. The observed strong drops of the 100-300 MeV flux indicate the variability 
of the total gamma-ray flux in the synchrotron cut-off regime. 
While the previously discovered giant gamma-ray flares could
be explained as a separate component over the quiescent synchrotron
gamma-ray flux, the flux drops we have found indicate the presence of coherent
variations of the whole source of the 100-300 MeV photons. The depressions we found are consistent with the complete lack of the synchrotron component above 100 MeV in the certain time intervals.
The apparent quasi-periodicity of the gamma-ray flux in 100-300 MeV range can be understood if the gamma-ray radiation is produced in the downstream of the termination surface which has reformation time scale (see e.g. Fig. 6 in \citet{Porth14}) consistent with the quasi-period. 

This may
provide important constrains for the models of synchrotron gamma-ray
radiation in PWNe. Indeed the models of impulsive acceleration 
of multi-PeV energy regime pairs by a reconnection event or a short pulse of Doppler
boosted gamma-rays from some localized regions can be responsible for
the flares, but on another hand, the strong drops of the gamma-ray flux can be
associated with variations of magnetic field within the source of the
synchrotron gamma-ray photons (e.g., the wind termination surface). 
We have shown that relatively modest variations of magnetic field can produce 
strong flux variations in the synchrotron cut-off regime \citep{Bykovea12}. 
In such a simple model the variable flux emerges near the PWN termination surface 
due to the inhomogeneities of the pulsar wind magnetic field inflowing 
into the quasi-steady distribution of the locally accelerated PeV energy regime pairs. This simple approach could explain both the strong flares and long depressions. Sharp gamma-ray flux drops of duration $\sim 2\times10^5$ s in the Crab Nebula models require the relativistic velocities of the  flows downstream of the termination surface. We estimate that the emission from the flow with $\Gamma_d\sim3$ could explain sharp drops of the order of $2\times10^5$ s, but this result is sensitive to the flow geometry  and future simulations of Crab PWN plasma flow are needed. 
Future observations of multi-TeV photons from the Crab nebula and especially observations of variability patterns in this energy range, would certainly help us to distinguish between different models of high energy emission of the nebula.

\section*{Acknowledgements}
The authors sincerely thank the anonymous referee for a careful reading of our paper, the very constructive comments and a suggestion to discuss the flux periodicity issue.  
MSP and BAN acknowledge support of Leading Science School MSU (Physics
of Stars, Relativistic Compact Objects and Galaxies) and support by the
Foundation for the Advancement of Theoretical Physics and Mathematics
``BASIS'' grant 18-1-2-51-1. This research has made use of NASA's 
Astrophysics Data System. The numerical part of the work was done at the computer cluster of the Theoretical Division of INR RAS and at the ``Tornado'' subsystem of the St. Petersburg Polytechnic University supercomputing center.  

\section*{Data availability}
 The data underlying this article were accessed from Fermi-LAT website (https://fermi.gsfc.nasa.gov/cgi-bin/ssc/LAT/LATDataQuery.cgi) and Jodrell Bank Observatory Crab pulsar monthly ephemeris webpage (http://www.jb.man.ac.uk/~pulsar/crab.html). The derived data generated in this research will be shared on reasonable request to the corresponding author. 

\bibliographystyle{mnras}


\end{document}